\documentclass[twocolumn,showpacs,preprintnumbers,amsmath,amssymb,superscriptaddress]{revtex4-1}

\usepackage{graphicx}
\usepackage{dcolumn}
\usepackage{bm}
\usepackage{color}
\usepackage{natbib}

\begin{document}

\title{Magnetization dynamics of a CrO$_2$ grain studied by
micro-Hall magnetometry} 

\author{P.\ Das}
 \email{das@physik.uni-frankfurt.de}
\affiliation{Institute of Physics, Goethe University, Max-von-Laue Str.\ 1, 60438 Frankfurt (M), Germany} 
\affiliation{Max Planck
Institute for Chemical Physics of Solids, 01187 Dresden, Germany}
\author{F.\ Porrati}
\affiliation{Institute of Physics, Goethe University, Max-von-Laue
Str.\ 1, 60438 Frankfurt (M), Germany}
\author{S.\ Wirth}
\affiliation{Max Planck Institute for Chemical Physics of Solids,
01187 Dresden, Germany}
\author{A.\ Bajpai}
\affiliation{IFW-Dresden, Institute of Solid State Research, 01171
Dresden, Germany}
\author{M.\ Huth}
\affiliation{Institute of Physics, Goethe University, Max-von-Laue
Str.\ 1, 60438 Frankfurt (M), Germany}
\author{Y.\ Ohno}
\affiliation{Laboratory for Nanoelectronics and Spintronics,
Research Institute of Electrical Communication, Tohoku University,
Sendai, Japan}
\author{H.\ Ohno}
\affiliation{Laboratory for Nanoelectronics and Spintronics,
Research Institute of Electrical Communication, Tohoku University,
Sendai, Japan}
\author{J.\ M\"uller}
 \email{j.mueller@physik.uni-frankfurt.de}
\affiliation{Institute of Physics, Goethe University, Max-von-Laue
Str.\ 1, 60438 Frankfurt (M), Germany}

\begin{abstract}
Micro-Hall magnetometry is employed to study the magnetization
dynamics of a single, micron-size CrO$_2$ grain. With this
technique we track the motion of a single domain wall, which
allows us to probe the distribution of imperfections throughout
the material. An external magnetic field along the grain's easy
magnetization direction induces magnetization reversal, giving rise to a series of sharp jumps in magnetization.
Supported by micromagnetic simulations, we identify the transition
to a state with a single cross-tie domain wall, where pinning/depinning of the wall results
in stochastic Barkhausen jumps.
\end{abstract}

\maketitle

Studying the behavior of small magnetic particles \cite{Barbara01}
is important both from the fundamental point of view -- for
testing basic concepts of ferromagnetism -- and for potential
applications in high-density magnetic storage, magnetic sensing,
spintronics, or in biology. In order to simplify the analysis
of magnetization behavior to an accessible level many
experimental studies have been performed on small arrays or even
on single nanometer-scale particles
\cite{Wernsdorfer1996,Dunin1999,Wirth1999,Li2002}. On the other
hand, calculations and numerical simulations have revealed pivotal
insight into possible modes of magnetization reversal, see e.g.\
\cite{Braun94,Hinzke00,Brown09}. For the magnetization
behavior of systems, which are larger than the critical size of
coherent magnetization reversal, domain walls (DWs) play a
decisive role. Again, it is often beneficial to study a
single entity, small enough that its magnetization dynamics is
governed by a single DW \cite{Novoselov2003, Christian2006}. One
reason is that DW motion may yield indications of the intrinsic
defect structure of a magnetic particle. The calculation of
magnetic key quantities, such as the coercivity and the remanence,
in turn requires knowledge of the nature of imperfections in the
material and how they contribute to the various energy terms, as
well as the
distribution of imperfections throughout the material. 

\begin{figure}[t]
\begin{center}
\includegraphics[width=0.475\textwidth]{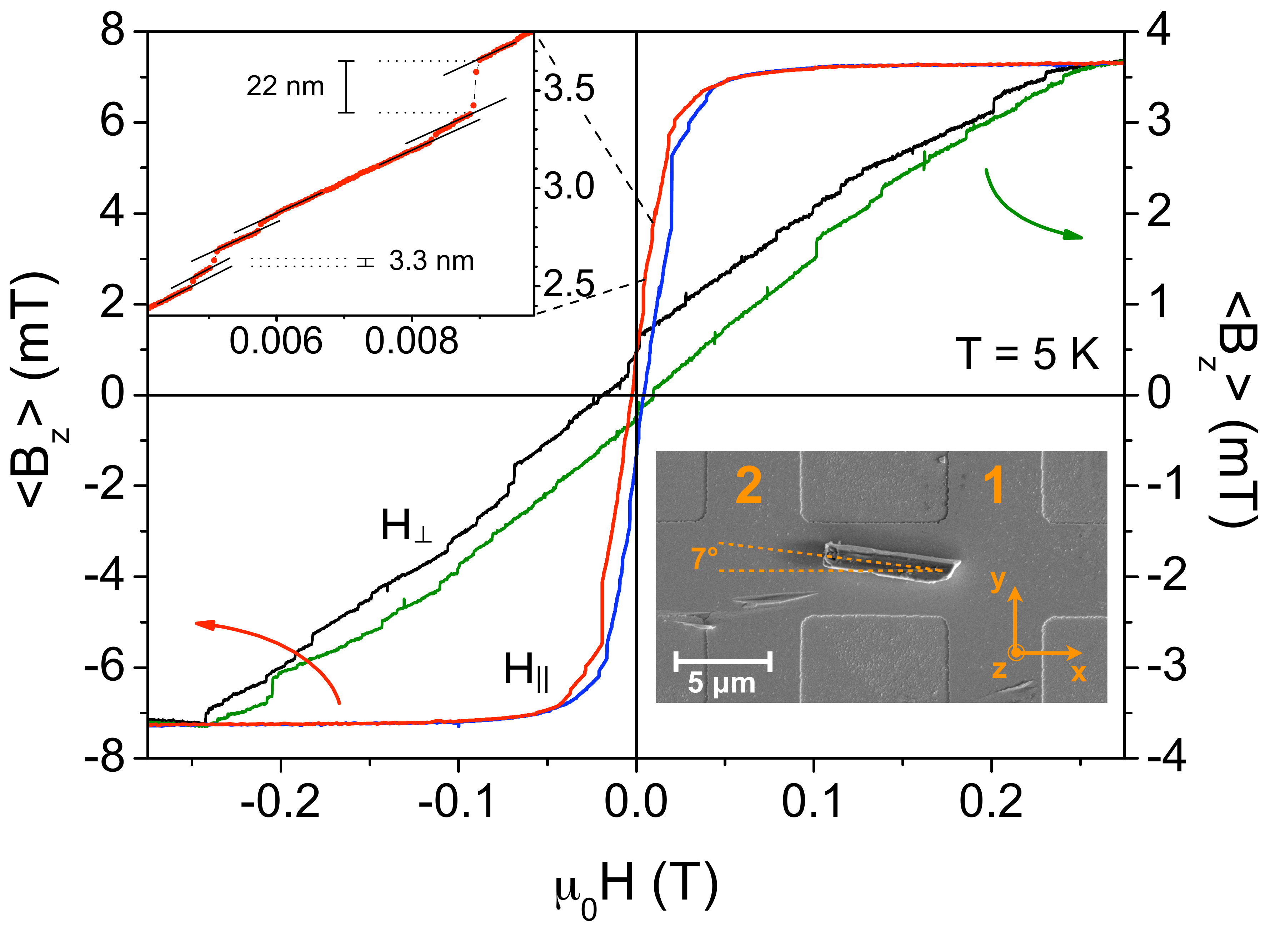}
\caption{\label{fig1} (Color online) Hysteresis loops measured for
magnetic fields applied almost 
parallel and perpendicular to the easy magnetization direction (EMD). 
Lower right inset: SEM image of the CrO$_2$ grain placed
between two Hall crosses (labeled 1 and 2). Upper left inset: Barkhausen jumps
corresponding to the displacement of a single domain wall (see
text).}
\end{center}
\end{figure}

Many of the previous single-wall studies were performed on either
permalloy or garnet films \cite{Novoselov2003, Christian2006,
Atkinson2003, Parkin2008}. In this Letter, we report results
of magnetization measurements of an individual, single-crystalline
CrO$_2$ grain of rod-like shape with dimensions of approximately
$(7 \times 1.2 \times 0.5)\,\mu$m$^3$, see inset of
Fig.\,\ref{fig1}. CrO$_2$ has been known as a non-volatile
magnetic-storage material for a long time and has attracted
renewed interest\cite{Coey02} as an almost completely
half-metallic ferromagnet ($T_{\rm C} \approx$ 393\,K) being
explored as a potential candidate for spintronics
applications.
For this material, in particular, the magnetotransport properties
of grains, nanowires and thin films being influenced by the
dynamics of magnetic DWs have been subject of recent
interest \cite{Coey1998,Koenig2007,Zou2008}. CrO$_2$ has the
tetragonal rutile structure and a uniaxial magnetocrystalline
anisotropy with the easy magnetization direction (EMD) along
the [001] direction \cite{Lewis1997,Goering2002}. This direction
coincides with the longest dimension of our rod such that
magnetocrystalline and shape anisotropy simply superpose.
High-purity crystallites for this study were grown in a
two-step process using a new route of synthesis that allows
to control the grain size and shape \cite{Bajpai2005}.

The motivation for our study was to gain a quantitative
understanding of the magnetization reversal of a {\em single}
CrO$_2$ grain. In micromagnetic simulations carried out for a
rod-like grain with rectangular surfaces and the dimensions of our
sample we find that the ground state (demagnetized state)
corresponds to a single cross-tie DW along the long axis of the
grain. Experimentally we observe that in the regime of single DW
motion, the magnetization reversal in the CrO$_2$ micro-grain
takes place through a series of sharp Barkhausen jumps, which we
attribute to stochastic pinning and depinning of the DW.
Accordingly, the magnetic flux densities emanating from both ends
of the grain are strictly correlated and change in opposite
directions.

For the magnetic measurements, we analyzed \cite{Wirth1999,Li2002}
the Hall response of a two-dimensional electron gas (2DEG) 
to the $z$-component of the stray field $B_z$ emanating from
a CrO$_2$ grain positioned on top of the 2DEG, {\em cf.}
inset of Fig.~\ref{fig1}. To allow for simultaneous
measurement of the stray fields at both ends of the grain, we use
Hall crosses of active area $(5 \times 5)$\,$\mu$m$^2$ whose
separation was chosen to correspond to the grain length.
The magnetic measurements were performed with external magnetic
field $\mu_0 H_{\rm ext}$ applied along the $x$- and
$y$-directions, {\it i.e.}\ almost parallel (EMD) and
perpendicular (hard axis) to the long axis of the
grain, respectively. 

Figure\,\ref{fig1} shows the hysteresis loops of the grain
measured at $T = 5$\,K for both directions of $H_{\rm ext}$.
The expected pronounced anisotropy is obvious.
For the easy-axis magnetization we find a relatively low coercive
field of $H_c \sim$ 3.5\,mT which is consistent with the values
reported for bulk and thin films~\cite{Bajpai2005, Li1999}.
Analyzing the magnetization curve measured along the magnetically
hard direction in terms of reversible magnetization rotation the
anisotropy field $\mu_0 H_a \sim 250$ mT can be obtained.
This value yields an effective uniaxial anisotropy
constant of $K_{\rm{eff}} = H_{a} J_s/2 \sim 7.3 \times 10^4$\,J/m$^3$, where
$J_s \approx$ 0.74 T is used for the spontaneous polarization at low
temperatures~\cite{Bajpai2010}.
The shape anisotropy constant $K_s = (N_\perp -
N_\parallel) J_s^2/2 \mu_0$ for the grain is found to be $\sim 6 \times 10^4$\,J/m$^3$. Therefore, the uniaxial magnetocrystalline anisotropy constant $K_1$ is $\sim\,1.3 \times 10^4$\,J/m$^3$
which is consistent with the values of $1.4 - 6 \times
10^4$\,J/m$^3$ reported in the literature
\cite{Zou2007, Koenig2007,Li1999,Rodbell1966,Goering2002}. 
For both field directions, we observe
that the magnetization changes through a series of sharp jumps.
Assuming the motion of a {\em single} DW for the 
EMD, see below, the change in the
average stray field $\langle B_z \rangle$ due to the DW displacement
$\Delta y$ within the grain is described by $\Delta \phi = 2 M_s d
\Delta y$, where $\phi \propto \langle B_z \rangle$ is the flux
emanating from the grain (approximated by a rod of thickness $d$)
\cite{Bertotti}. The upper left inset of
Fig.~\ref{fig1} indicates that the displacement of the
DW occurs in discrete steps $\Delta y$ of order of a few nm. The
smallest step we have observed corresponds to $\Delta y \approx
0.8$\,nm, demonstrating the high resolution of the present
magnetization measurement technique.

\begin{figure}[t]
\includegraphics[width=0.45\textwidth]{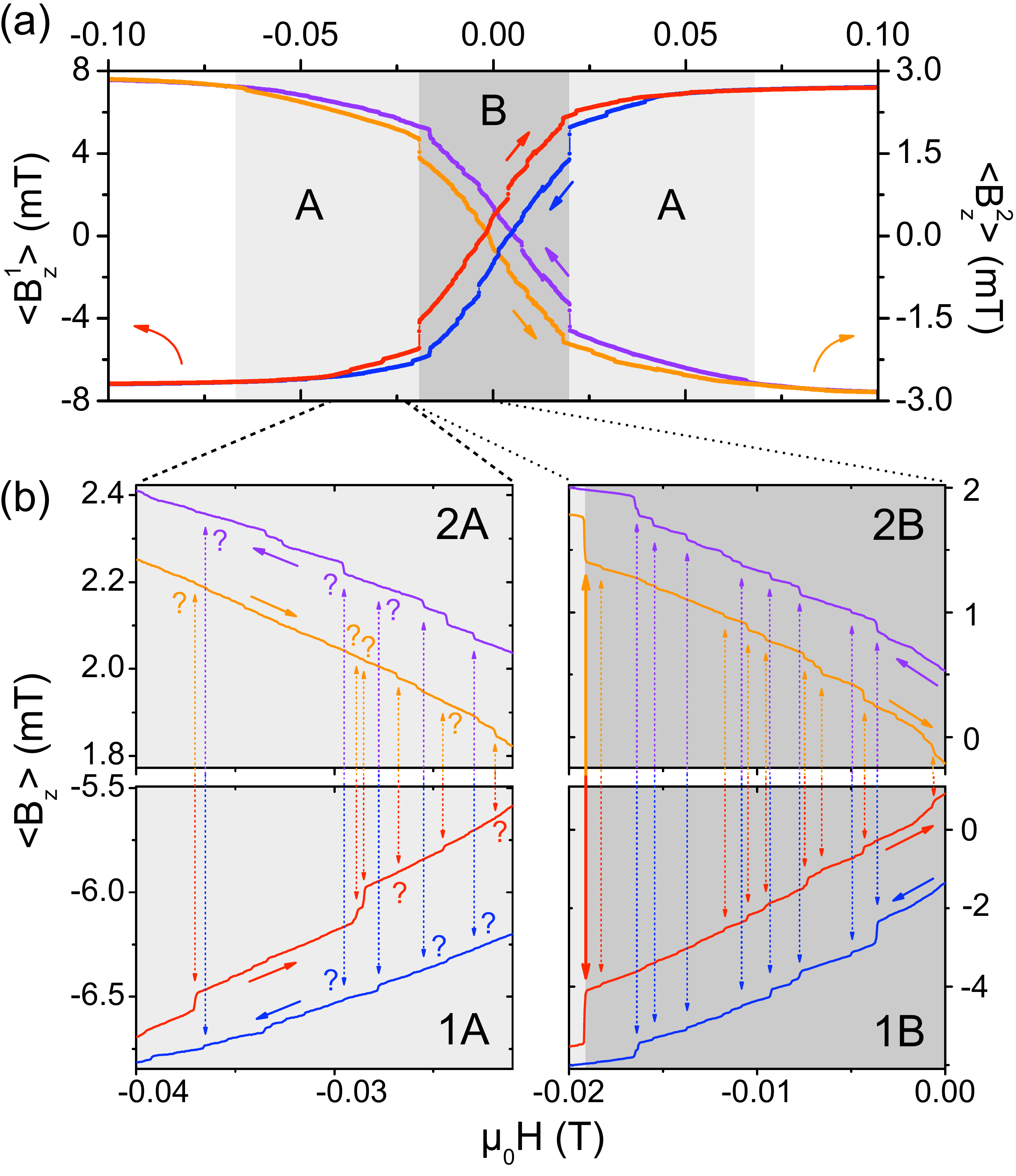}
\caption{\label{fig2} (Color online) (a) Easy-axis hysteresis loops simultaneously
measured at two grain ends (crosses 1 and 2) at $T = 5$\,K.
The sweep direction is indicated by the arrows. The apparent
negative remanence (see also Fig.~\ref{fig1}) is due to the
specific nature of the stray field $\langle B_z \rangle$ to
which our Hall crosses respond. The different regimes A and B
of magnetization reversal (indicated by the shaded areas)
correspond to different domain configurations, see text and
Fig.~\ref{fig3}. (b) Magnification of the hysteresis loops in
regions A ($|\mu_0H_{\rm ext}| > 20$\,mT) and B
($|\mu_0H_{\rm ext}| \leq 20$\,mT).
The jumps at the grain ends are strictly correlated within
regime B, {\it i.e.}, they appear at exactly the same field
values for identical sweep directions (panels 1B and 2B). In
contrast, jumps are mostly uncorrelated for regime A (panels 1A
and 2A, jumps without counterpart are highlighted by question
marks).}
\end{figure}
Figure~\ref{fig2} shows the hysteresis loops measured at both ends
of the grain for the external magnetic field along the EMD. The
magnetic signal at the two ends are oppositely directed. For the
hysteresis loops, we identify two regimes of magnetization
behavior: below and above $\mu_0 H^\ast \approx \pm 20$\,mT, where
a large jump in magnetization
occurs.\ 
The transition field $H^\ast$ is not stochastic and
essentially temperature independent up to the highest temperature
of our experiment of 100\,K. All other discontinuous magnetization
changes occur at random, {\it i.e.}, in a stochastic manner.

For $|H_{\rm ext}| < H^\ast$, the magnetization changes
according to Barkhausen jumps which are of opposite sign and
strictly {\em correlated} for the flux through crosses 1 and 2.
This correlation is clearly seen in Fig.~\ref{fig2}(b), panels 1B
and 2B, where the identical field values at which jumps are
detected from the two ends of the grain are highlighted by
vertical arrows.
We attribute this part of the hysteresis curve to
pinning/depinning of a single DW. Beyond the large jump, {\it
i.e.}\ for $|H_{\rm ext}| > H^\ast$ (panels 1A and 2A of
Fig.\,\ref{fig2}(b)), the magnetization changes through rotation
of magnetic domains and Barkhausen jumps, which are mostly {\em
uncorrelated} for both ends of the grain.

\begin{figure}[t]
\begin{center}
\includegraphics[width=0.45\textwidth]{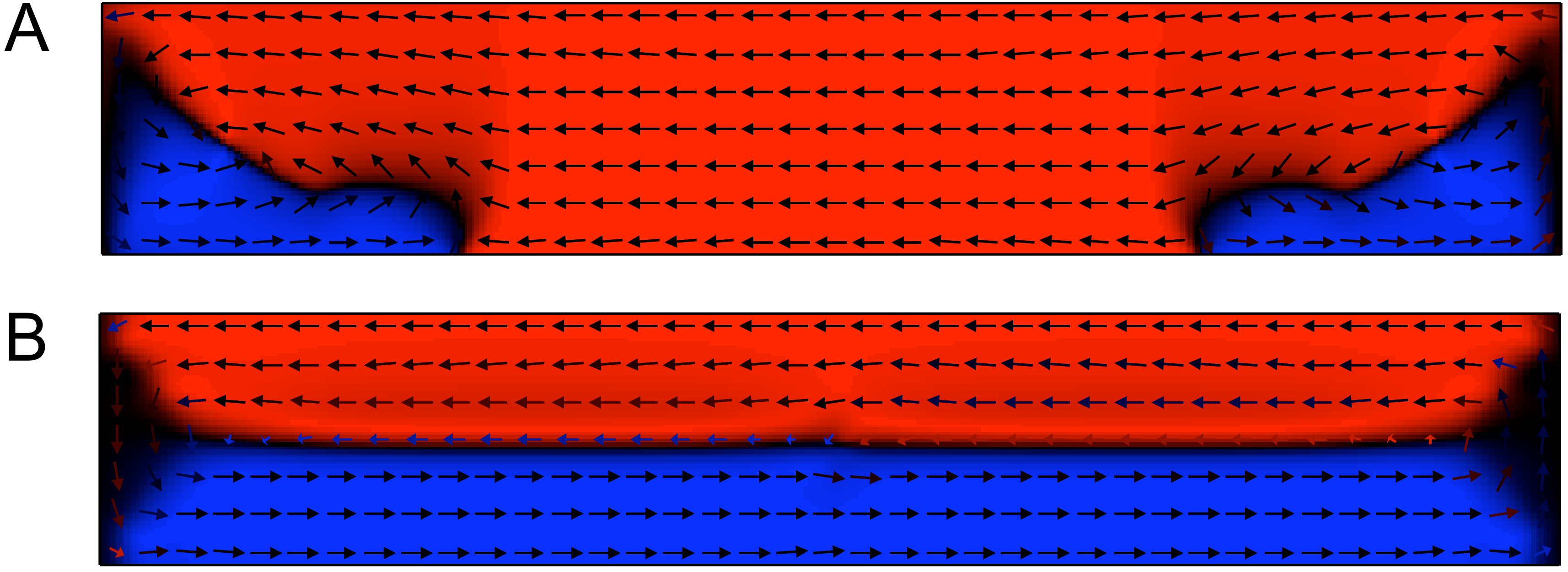}
\caption{\label{fig3}(Color online) Results of
micromagnetic simulations as described in the text. The figures
show one of the two surfaces of the grain in the $x$-$y$-plane. One
arrow corresponds to approximately 6 unit cells of discretization.
The different domain configurations A and B correspond to the
different magnetization regimes shown in Fig.\,\ref{fig2}.}
\end{center}
\end{figure}
Micromagnetic simulations~\cite{comment3} support the conjecture of the
existence of these regimes which are related to two distinct
states, correspondingly labeled A and B in Fig.~\ref{fig3}. 
For the simulations an exchange stiffness constant of
$A\,=\,4.6\,\times\,10^{-12}$ J/m has been used
\cite{Zou2007} together with the anisotropy constant $K_1$ and the spontaneous magnetization $J_s$ as mentioned above.
The simulation volume is discretized in unit cells of
cubic shape with an edge length of 18 nm.
The magnetization configurations are obtained in the quasistatic
regime after relaxation of the initial configuration.\
The configuration B is obtained from a two-domain
configuration with magnetization oppositely oriented along
the easy axis. During relaxation the magnetic moments close to
the wall between the two domains curl forming a cross-tie wall.
Alternatively, configuration B can be obtained from A by applying
a small field anti-parallel to the main magnetic domain. Configuration B is the ground state of the system, with a total energy about two third of configuration A. Configuration A is obtained after relaxation of a three-domain configuration imposed as initial state. A similar state to configuration A is attained by relaxation of the single
domain state with the magnetization oriented along the easy axis.
During the relaxation the magnetic moments close to the edges of
the sample curl in order to decrease the magnetostatic energy. 
The micromagnetic simulations suggest that the large jump at $\pm H^\ast$ marks
the transition from the metastable state A to ground state B,
{\it i.e.}\ the occurrence of the cross-tie DW. The magnitude
of $H^\ast$ depends on $J_s$ and $A$, both of which are
only weakly temperature dependent for $T < 100$ K because of the
comparatively high $T_{\rm C}$.\\
At fields $|H_{\rm ext}| > H^\ast$, the magnetization changes
may occur independently at both ends of the crosses, governed by
rotation and (de)pinning of the magnetic domains related to the
endcaps of the grain. Accordingly, as shown in Fig.~\ref{fig2}(a),
the opening of the hysteresis loop does not necessarily take
place at exactly the same fields for both ends and opposite field
directions.

In ferromagnetic materials of dimensions larger than the critical
single domain size, coercive mechanisms are domain nucleation
and pinning/depinning during DW motion. The material parameters
of the present CrO$_2$ grain favor comparatively strong pinning
forces, so the latter mechanism is of importance. We suggest that
the jumps in magnetization that we observe in the intermediate
field regime (B) are due to the interaction of DWs with
crystal imperfections through modified magnetostatic,
anisotropy and/or exchange interactions, where a complicated
energy landscape containing a number of local minima and saddle
points is involved. A domain configuration (here: a single DW),
when pinned to one of these local minima, is depinned by the
application of $H_{\rm ext}$ which leads to the observed
Barkhausen jumps.
The synchronous magnetization changes at both ends are also
found for the hard axis magnetization (not shown) and for
temperatures $T \leq 100$\,K, whereas both the number of
Barkhausen jumps and their magnitude decrease with increasing
temperature, indicating a thermal contribution to the (de)pinning.
A detailed statistics of the Barkhausen jumps depending on
temperature and magnetic field orientation will be published
elsewhere. The finite slope between the Barkhausen discontinuities
observed in our experiment may suggests a significant elastic
bending of the DW, or likewise, reversible rotation within the
grain and/or growth of the endcaps.
From the number of Barkhausen jumps occurring in the specific
field interval at low temperatures we estimate a density of
pinning centers (of a certain strength sufficient to be detected
within our resolution) of $1 - 10\;\mu$m$^{-3}$, which implies a
high purity of the investigated grain.

The work was supported by the Deutsche Forschungsgemeinschaft (DFG) through the Emmy Noether program. 
A.B.\ acknowledges support through EU Marie Curie IIF project 040127-NAWMATCR.
We acknowledge the help of F.\ Wolny at IFW Dresden with
the placement of the CrO$_2$ grain on the Hall device.


\end{document}